\documentclass[%
 reprint,
 amsmath,amssymb,
 prl,
]{revtex4-2}

\usepackage{graphicx}
\usepackage{dcolumn}
\usepackage{bm}

\usepackage{hyphenat}

\usepackage{chngcntr}
\usepackage{xcolor}
\begin{document}
	
	
\title{Simulating nonnative cubic interactions on noisy quantum machines}
	
\author{Yuan Shi}
\email{shi9@llnl.gov}

\author{Alessandro R. Castelli}
\author{Xian Wu}
\author{Ilon Joseph}
\author{Vasily Geyko}
\author{Frank R. Graziani}
\author{Stephen B. Libby}
\author{Jeffrey B. Parker}
\author{Yaniv J. Rosen}
\author{Luis A. Martinez} 
\author{Jonathan L DuBois}	

\affiliation{Lawrence Livermore National Laboratory, Livermore, California 94551, USA}

\date{\today}
	
\begin{abstract}
	As a milestone for general-purpose computing machines, we demonstrate that quantum processors can be programmed to efficiently simulate dynamics that are not native to the hardware. Moreover, on noisy devices without error correction, we show that simulation results are significantly improved when the quantum program is compiled using modular gates instead of a restricted set of standard gates. 
	We demonstrate the general methodology by solving a cubic interaction problem, which appears in nonlinear optics, gauge theories, as well as plasma and fluid dynamics. 	
	To encode the nonnative Hamiltonian evolution, we decompose the Hilbert space into a direct sum of invariant subspaces in which the nonlinear problem is mapped to a finite-dimensional Hamiltonian simulation problem. 
    In a three-states example, the resultant unitary evolution is realized by a product of $\sim$20 standard gates, using which $\sim$10 simulation steps can be carried out on state-of-the-art quantum hardware before results are corrupted by decoherence. 
    In comparison, the simulation depth is improved by more than an order of magnitude when the unitary evolution is realized as a single cubic gate, which is compiled directly using optimal control. Alternatively, parametric gates may also be compiled by interpolating control pulses. 
    Modular gates thus obtained provide high-fidelity building blocks for quantum Hamiltonian simulations. 
\end{abstract}
	
\maketitle
	

Ideal quantum computers can in principle perform arbitrary unitary operations. However, a lengthy sequence of standard gates is required when the unitary operation results from Hamiltonian evolution that is not native to the hardware. In this case, implementing the operation becomes infeasible when the gate depth exceeds the coherence limit of noisy devices. 
While this issue may ultimately be resolved by quantum error correction \cite{Cory98,Chiaverini2004realization,Terhal15,Ofek2016extending,Reiter2017dissipative,Rosenblum18}, in this letter, we show that efficient compilation plays a key role in realizing quantum computing.
When compiling a quantum program using modular gates \cite{Shi2019optimized}, the end performance of quantum hardware is significantly improved due to the reduction of gate depth. 
In particular, we show that for cubic interactions, which are essential for nonlinear optics \cite{Bloembergen1996nonlinear}, gauge theories \cite{Yang1954conservation}, as well as plasma and fluid dynamics \cite{Davidson72}, modular gates enable useful Hamiltonian simulations on present-day noisy hardware, which usually lacks native cubic couplings \cite{Haffner2008quantum,Clarke2008superconducting}.

To close a major gap in the literature, we consider the following underexamined cubic Hamiltonian (\mbox{$\hbar=1$}) that governs the lowest-order nonlinear interactions in quantum optics and plasma dynamics:
\begin{equation}
\label{eq:H3}
H=ig A_1^\dagger A_2A_3 -ig^* A_1 A_2^\dagger A_3^\dagger,
\end{equation}
where $g$ is the coupling coefficient and $[A_j, A_l^\dagger]=\delta_{jl}$. The resulting Heisenberg equations are special cases of three-wave equations
\begin{eqnarray}
\label{eq:dA1}
d_tA_1&=&g A_2 A_3,\\
\label{eq:dA2}
d_tA_2&=&-g^*A_1 A_3^\dagger,\\
\label{eq:dA3}
d_tA_3&=&-g^*A_1 A_2^\dagger,
\end{eqnarray}
where $d_t=\partial_t+\mathbf{v}_j\cdot\nabla$ is the convective derivative at the wave group velocity $\mathbf{v}_j=\partial\omega_j/\partial\mathbf{k}_j$. 
As example applications, Eqs.~(\ref{eq:dA1})-(\ref{eq:dA3}) are solved to (i) optimize the input seed pulse shape when designing nonlinear optical systems in the pump depletion regime \cite{Frantz1963theory,Ahn2003terahertz,Brunton2012shaping}; 
(ii) determine the turbulence spectrum in the wave-kinetic approach to weak turbulence \cite{Zakharov2012kolmogorov}, 
and (iii) compute the energy delivered to fusion fuel when drive lasers exchange energy in plasmas \cite{Moody2012multistep, Myatt2014multiple, Shi2017three}.
More broadly speaking, cubic terms are common for nonlinear problems, for which cubic gates that enacts Eq.~(\ref{eq:H3}) serve as basic building blocks. Cubic gates, in conjunction with quartic gates that are usually natively available on hardware, allow one to simulate a majority of problems in physics.

In the usual approach to quantum \textit{emulation}, terms in the hardware Hamiltonian $H_0=\sum_{k=1}^m H_k$ are turned on or off to generate the unitary evolution due a targeted $H=\sum_{k=1}^m a_k H_k$, where $a_k$ are real coefficients. Explicitly, using the 
Lie\hyp{}Trotter\hyp{}Suzuki 
expansion, $\exp(-iHT)=\lim_{n\rightarrow\infty}[\prod_{k=1}^m U_k (a_k T/n)]^n$, where $U_k (t)=\exp(-iH_k t)$. 
Difficulty arises when $H$ contains terms that are not natively available in $H_0$. 
Although cubic couplings may be realized natively with special-purpose hardware \cite{Bergeal2010phase,Bergeal2010analog}, direct emulations have two major limitations. 
First, qubits can only be in a superposition of $|0\rangle$ and $|1\rangle$ states, whereas a large amplitude wave, such as that of a laser, involves many photons. Therefore, directly representing wave states by qubit states is inefficient. 
Second, parameters of qubits may not be easily adjustable \cite{Chen14} to generate Eq.~(\ref{eq:H3}). Since quantum hardware likely contains additional terms in $H_0$ that cannot be completely turned off, direct emulation is demanding of the hardware tunability and may not be realizable for problems of practical interest.

In comparison, \textit{simulation} that encodes the interaction at software level with reusable modular gates is more versatile and allows quantum processors to operate as general-purpose computing machines, which do not need to be rebuilt to solve a different problem.
In what follows, we construct an efficient algorithm in the action space to enact the nonnative cubic Hamiltonian evolution. Here, the term ``action" originates from the action-angle variables of the slow-envelope approximation, which is commonly used to derive the three-wave equations.
For Eq.~(\ref{eq:H3}), two independent action operators that commute with $H$ are
\begin{eqnarray}
S_2=n_1+n_3,\\
S_3=n_1+n_2,
\end{eqnarray}
where $n_j= A_j^\dagger A_j$ is the number operator. The simultaneous eigenspace of $H$, $S_2$ and $S_3$ has dimension $D=\min(s_2,s_3)+1$, where the integer $s_j\ge0$ is the eigenvalue of $S_j$. Without loss of generality, suppose $s_2\leq s_3$, which breaks the $2\leftrightarrow 3$ symmetry. 
Then, in the Fock basis $|n_1,n_2,n_3\rangle$, any state in the $D$-dimensional subspace 
$V$ is spanned by
\begin{equation}
|\psi\rangle=\sum_{j=0}^{s_2} \alpha_j |s_2-j,s_3-s_2+j,j\rangle,
\end{equation}
where the expansion coefficients satisfy the normalization condition $\sum_j |\alpha_j|^2=1$. It is easy to check that $S_j|\psi\rangle=s_j|\psi\rangle$. 
Moreover, it is important to recognize that $V$ is a closed subspace under the action of $H$, namely, $H|\psi\rangle\in V$ whenever $|\psi\rangle\in V$. Therefore, the total Hilbert space can be decomposed into a direct sum of invariant subspaces, and the Hamiltonian dynamics is confined within the subspaces specified by initial conditions. 
Notice that using $D=s_2+1$ levels, the above mapping can efficiently represent $s_3\gg s_2$ photons.

Moreover, using the action-space algorithm, the nonlinear three-wave problem is mapped to a finite-dimensional linear Hamiltonian simulation problem. In the Schr\"{o}dinger picture, $i\partial_t|\psi\rangle=H|\psi\rangle$ becomes a system of equations for $\alpha_j(t)$
\begin{eqnarray}
i \partial_t \alpha_j
=igh_{j+\frac{1}{2}}\alpha_{j+1}-ig^*h_{j-\frac{1}{2}}\alpha_{j-1},
\end{eqnarray}
where $h_{j-\frac{1}{2}}=\sqrt{j(s_2+1-j)(s_3-s_2+j)}$ with $h_{-\frac{1}{2}}=h_{D+\frac{1}{2}}=0$.
In other words, $H$ is represented by a block tridiagonal matrix with zero diagonal elements.
The above mapping utilizes the special form of $H$, but similar algorithms may be constructed for other nonlinear problems when the Hamiltonian has special symmetries.

Once the expansion coefficients are solved for given initial conditions, observables of interest are evaluated by $O(D)$ classical operations during post processing. For example, occupation numbers of the three waves are given by $\langle n_1 \rangle=\sum_{j=0}^{s_2} (s_2-j) |\alpha_j|^2$, $\langle n_2 \rangle=\sum_{j=0}^{s_2} (s_3-s_2+j) |\alpha_j|^2$, and $\langle n_3 \rangle=\sum_{j=0}^{s_2} j |\alpha_j|^2$.
Higher-order cumulants, thereby all possible expectation values of interest, can be obtained similarly, and the quantum three-wave problem is then solved. 

Before solving the Schr\"{o}dinger equation, it is instructive to also analyze the cubic interactions in the Heisenberg picture. The number operators satisfy
\begin{eqnarray}
\partial_t^2 n_1
&=&-\partial_t^2 n_2=-\partial_t^2 n_3 \\
\nonumber
&=& 2|g|^2\big[s_2s_3-(2s_2+2s_3+1) n_1+3n_1^2\big].
\end{eqnarray}
In comparison, the classical wave action $I_j = |A_j|^2$,
where $A_j$ is treated as a complex number, satisfies
\begin{eqnarray}
\partial_t^2 I_1&=&-\partial_t^2 I_2=-\partial_t^2 I_3 \\
\nonumber
&=&2|g|^2\big[s_2s_3-(2s_2+2s_3) I_1+3I_1^2\big].
\end{eqnarray}
When identifying $I_j\simeq\langle n_j\rangle$, the difference between quantum and classical systems is proportional to $3(\langle n_1^2\rangle-\langle n_1\rangle^2)-\langle n_1\rangle$. 
The term in the parenthesis
is exactly zero for states with Poisson statistics, and is usually small compared to other terms when the state is a well-localized semi-classical state. 
The second term is comparatively small when $s_2$ and $s_3$ are large, whereby stimulated processes, which depend quadratically on the number of photons, dominate spontaneous emission, which depends linearly on the number of photons.
When both terms are small, the system is in the semiclassical regime \cite{Jaynes63}, where the quantum and classical solutions are similar. In the opposite regime, the behaviors of the two systems are disparate. For example, in the thermodynamic limit where occupation numbers satisfy the Boltzmann distribution $|\alpha_j|^2=1/D$, the quantum system is stationary, whereas the classical system, which is solved by Jacobi elliptic functions \cite{Jurkus1960saturation,Armstrong62}, is nonstationary.

In the following, we demonstrate the algorithm using three levels, or equivalently, the $|00\rangle$, $|01\rangle$, and $|10\rangle$ states of two qubits. The allowable dimension is $D\le3$, and we take $s_2=2$ and $s_3=s\ge2$. Then, a natural mapping 
is $|2,s-2,0\rangle=(1,0,0)^\text{T}$, $|1,s-1,1\rangle=(0,1,0)^\text{T}$, and $|0,s,2\rangle=(0,0,1)^\text{T}$.
The normalized Hamiltonian $h=H/|g|$, when restricted to the invariant subspace, is 
\begin{equation}
\label{eq:h3}
h(\theta, s)=\left( \begin{array}{ccc}
0 & e^{i\theta}\sqrt{2(s-1)} & 0 \\
e^{-i\theta}\sqrt{2(s-1)} & 0 & e^{i\theta}\sqrt{2s} \\
0 & e^{-i\theta}\sqrt{2s} & 0
\end{array} \right),
\end{equation}
where $\exp(i\theta)=ig/|g|$.
It is worth emphasizing that the general case is associated with a $D\times D$ matrix, where $D\simeq2^n$ with $n$ being the number of qubits needed. In other words, $D$ needs not be three, but the Hamiltonian matrix is always tridiagonal.  
For the three-level problem, Eq.~(\ref{eq:h3}) is exponentiated analytically \footnote{See Supplemental Material for the unitary matrix of the three-level problem.} 
to determine the unitary time-evolution operator $U(\Delta\tau, \theta, s)=\exp[-ih(\theta, s)\Delta\tau]$. 
This unitary matrix, for a given $\Delta\tau=|g|\Delta t$, is an input for quantum compilers. 
Notice that the quantum hardware may not be able to track $U(\tau)$ continuously, because the cubic Hamiltonian has odd parity whereas the native Hamiltonian may not be. Nevertheless, at discrete time steps, $U(\Delta\tau)$ can be realized just like other complex unitary operators.

\begin{figure}[t]
	\begin{center}
		\includegraphics[width=0.45\textwidth]{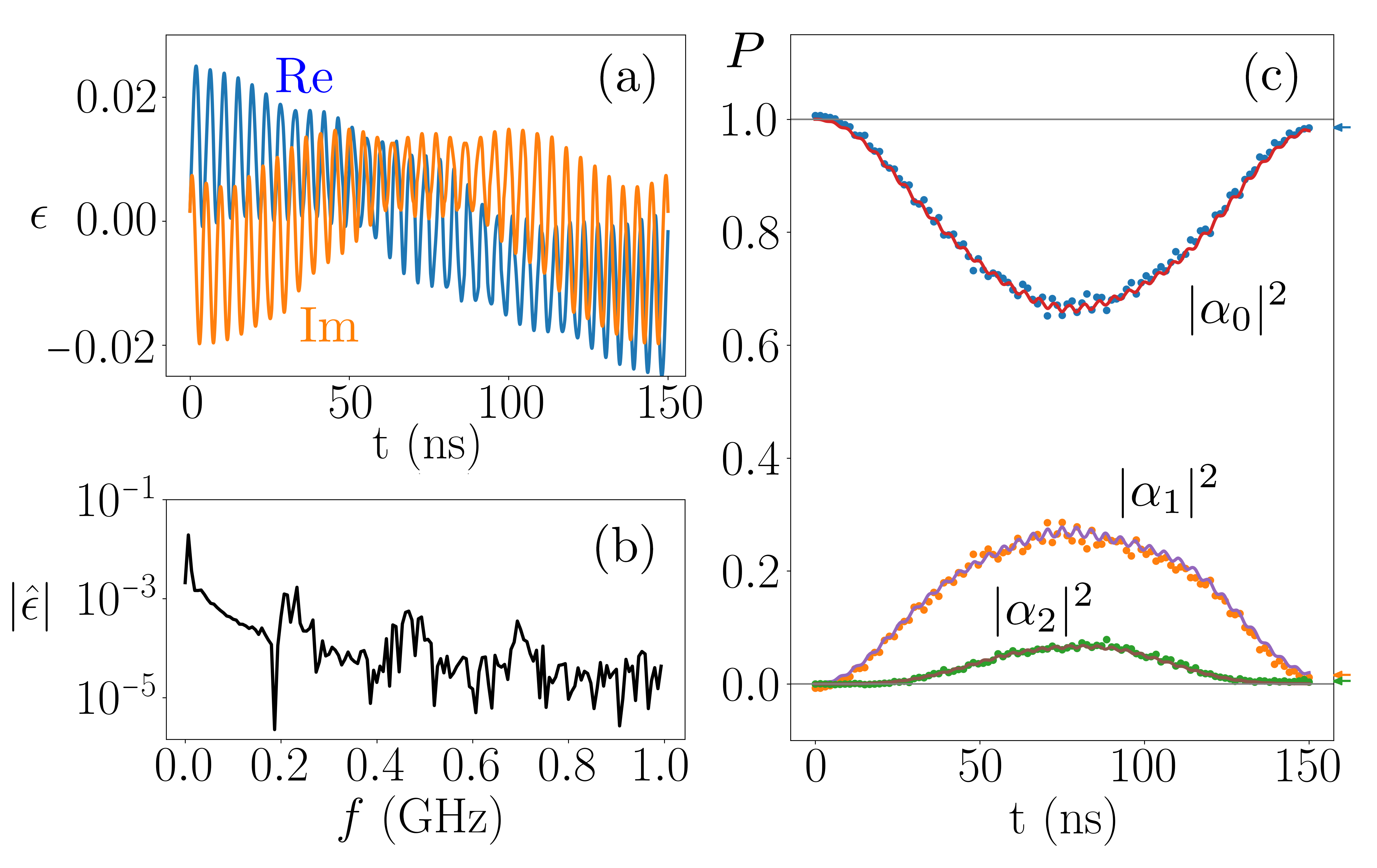} 
		\caption{(a) Optimized control pulse $\epsilon(t)$ for achieving the unitary operator $U(0.1, \pi/2, 2)$ on a transmon qudit. (b) The Fourier transform $\hat{\epsilon}(f)$ shows low-frequency modulations and peaks at harmonics of $f_{01}-f_{12}$\mbox{$\approx 0.23$ GHz}, where $f_{ij}$ is the qudit $|i\rangle\leftrightarrow |j\rangle$ transition frequency. (c) Experimentally estimated occupations $|\alpha_j|^2$ (dots) during the control pulse application match solutions to the master equation (lines). At the end of the pulse application, the targeted transition probabilities $P$ (arrows) are attained.
		}
		\label{fig:Control}
	\end{center}
\end{figure}

To compile the three-level unitary operator into a hardware executable form, the standard approach approximates it using a sequence of universal gates. 
As an example of state-of-the-art quantum cloud services, we implement the unitary operator on Aspen-4-2Q-A of Rigetti Computing \cite{Karalekas2020quantum,Smith2016practical}. The probabilistic Quil compiler converts $U(\Delta\tau, \theta, s)$ to a sequence of $\sim20$ native gates, including single-qubit Pauli gates and two \textsf{CZ} gates 
\footnote{See Supplemental Material for an example the Quil program.}. 

As a more hardware-efficient compilation approach, we realize the unitary operator as a single gate using optimal control: Instead of preparing control pulses for standard gates and then using these gates to approximate a targeted operation, we directly optimize a control pulse for the targeted operation.
The device we use is a transmon placed inside a 3D superconducting aluminum microwave cavity, whose details are described in \cite{Wu20PRL}. In addition to enhancing the coherence time, this many-qudit single-readout architecture reduces the number of control lines required to access larger computational spaces, and allows simultaneous control of multiple subsystems without relying on their crosstalk \cite{Blais04,Wallraff2004strong,Koch07,Schreier08,Paik11,Rigetti12}. 
The control Hamiltonian is \cite{Gerry2005introductory}
\begin{equation}
\label{eq:Hd}
H_c(t) \simeq \sum_j (c_j + c^\dagger_j)\Big(\epsilon_je^{-i\Omega_j t} + \epsilon_j^*e^{i\Omega_j t}\Big),
\end{equation}
where the complex-valued $\epsilon_j(t)$ is the slowly-varying envelope of a microwave field whose carrier frequency is $\Omega_j$, and $c_j$ is its control operator 
\footnote{See Supplemental Material for an explicit expression of the control operator.}. 
The time dependence of $\epsilon_j(t)$ may be engineered such that 
\begin{equation}
\label{eq:analog}
\mathcal{T} e^{-i\int_0^T dt [H_0+H_c(t)]}\simeq e^{-iH \Delta t},
\end{equation}
where $\mathcal{T}$ is the time-ordering operator, $T$ is the length of the control pulse, and $H_0$ is the hardware Hamiltonian. 
The control is not unique but can be implemented using direct digital synthesis \cite{Raftery2017direct,Heeres2017implementing} and designed using numerical optimization \cite{Khaneja05,Fouquieres2011second,Lloyd2014control,Petersson2020discrete} for small problems. 
For larger problems, the tridiagonal Hamiltonian may be decomposed as $H=\sum_{k=1}^m H_k$, where each $H_k$ is block diagonal. Each block in $H_k$, which is small enough and decoupled from other blocks at the hardware level, can now be enacted using numerically optimized control pulse prepared for the subsystem. Subsequently, the full Hamiltonian evolution due to $H$ can be approximated using the 
Lie\hyp{}Trotter\hyp{}Suzuki 
expansion.

As a test case, we construct a control pulse to achieve $U(0.1, \pi/2, 2)$. 
Based on a thorough characterization that provides parameters of the native Hamiltonian \cite{Wu20PRL}, the control pulse is generated in the $|0\rangle\leftrightarrow |1\rangle$ rotating frame using \textsf{optimize\_pulse\_unitary} in QuTIP \cite{Johansson13, Johansson2012qutip}, which calls an optimizer that uses the gradient ascent algorithms to iteratively update the control pulse to minimize the fidelity error.
The optimization uses zero as the initial guess, and is constrained by $|\epsilon|<0.03$. Two forbidden levels beyond the $D=3$ levels are included in the optimization to suppress possible state leakages at the end of the control pulse. The pulse duration (\mbox{$T=150$ ns}) is much shorter than the coherence time but long enough to allow the lowest three levels of the transmon, which has an anharmonicity of \mbox{$\sim0.23$ GHz}, to be separately addressed. 
The control pulse $\epsilon(t)$ is shown in Fig.~\ref{fig:Control}(a), and the absolute value of its Fourier transform $\hat{\epsilon}(f)$ is shown in Fig.~\ref{fig:Control}(b).  
The optimization takes $\sim 10$ iterations to reach convergence with an error target of $10^{-5}$.

When applying to the hardware, optimized control pulses are (i) concatenated in the rotating frame, (ii) transformed to the lab frame by mixing with a carrier wave at the qudit $|0\rangle\leftrightarrow|1\rangle$ transition frequency \mbox{$f_{01}\approx4.1$ GHz}, and (iii) compensated for spectral filtering effects of the hardware \cite{Wu20PRL}.
The final waveform is synthesized using an arbitrary waveform generator at \mbox{32-GHz} sampling rate, and is sent to the qudit inside a dilution refrigerator through a series of attenuators and a band-pass filter.
Occupations of the three levels are measured using dispersive readout \cite{Blais04,Boissonneault09,Wu20PRL}. The measured in-phase and quadrature signals are used to classify the states of the qudit, and the classification errors are partially corrected using a confusion matrix \cite{Wu20PRL}, which may cause unphysical occupations slightly above 1 or below 0.
The estimated occupations during the pulse application are shown in Fig.~\ref{fig:Control}(c) when the qudit is initialized in the ground state, which has minimal state preparation error. 
By taking into account hardware-specific decay and dephasing
\footnote{See Supplemental Material for the noise model based on the Lindblad master equation.}. 
the measurement results (dots) are well explained by numerical solutions (lines) of the Lindblad master equation \cite{Louisell1973quantum,Lindblad1976generators,Gorini1976completely,Schlosshauer2019quantum}. 
At the end of the control pulse, the intended unitary operator 
is realized, and the targeted transition probabilities (arrows) are attained. 
Using a modified process tomography \cite{Wu20PRL}, we estimate the process matrix of the cubic gate, which yields an average gate fidelity of 99.3\%.

\begin{figure}[t]
	\begin{center}
		\includegraphics[width=0.45\textwidth]{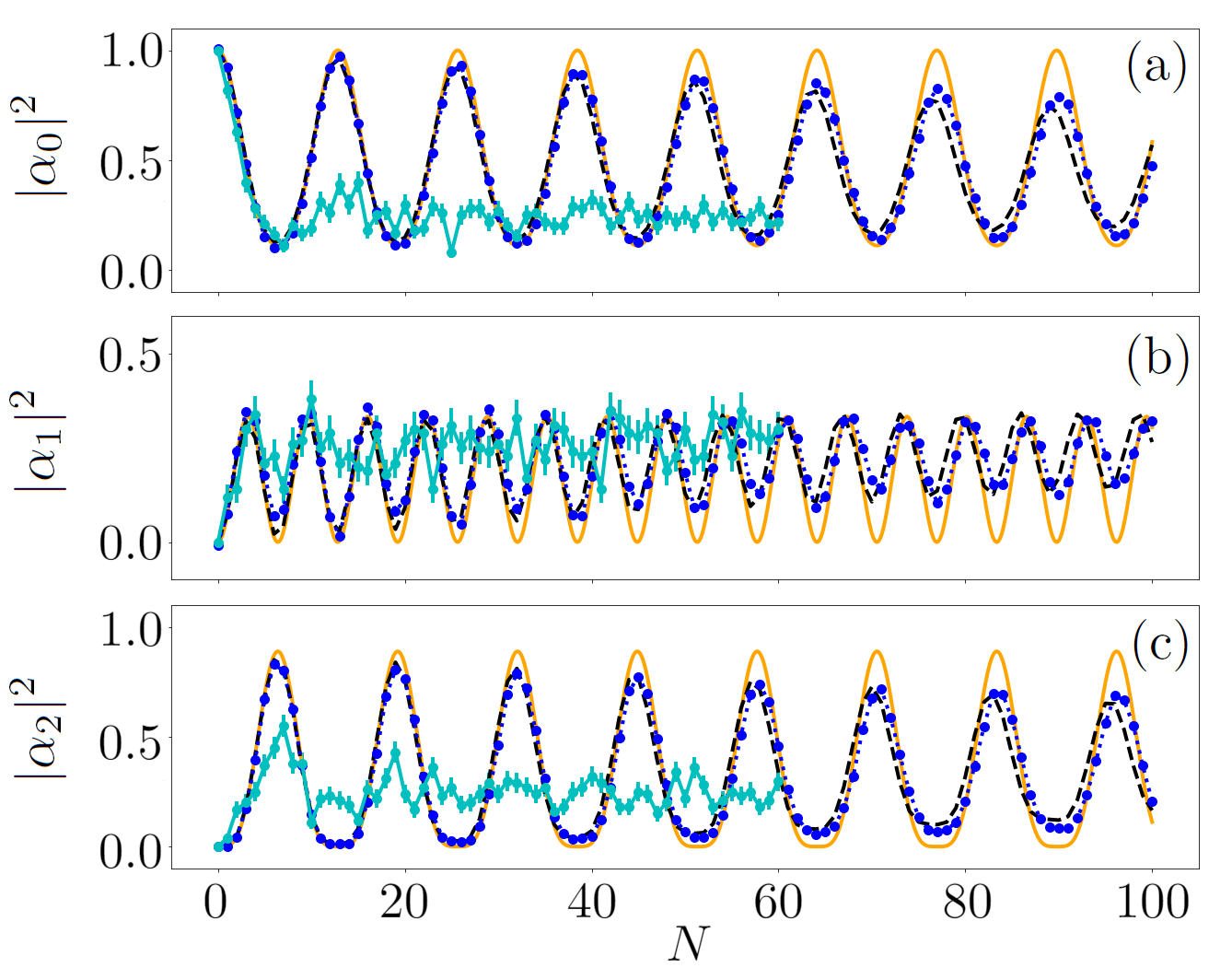} 
		\caption{Occupations of $|0\rangle$, $|1\rangle$, and $|2\rangle$ states
			after repeating $U(0.2, \pi/2, 2)$ for $N$ times. 
			By compiling $U$ as a single gate (blue), the simulation depth is improved by more than an order of magnitude compared to the standard compilation approach (cyan), which approximates each $U$ by $\sim 20$ native gates. Deviations from the exact results (orange) are explained by a noisy model based on the Lindblad master equation (black).}
		\label{fig:Results}
	\end{center}
\end{figure}

To compare the performance of the two compilation approaches, we repeatedly apply the unitary operator to simulate Hamiltonian time evolution. 
With the quantum processors initialized in the ground state, we read out occupations after repeating $U(0.2, \pi/2, 2)$ for $N$ times, and compare the results with the exact solutions (Fig.~\ref{fig:Results}, orange).  
Using a modular cubic gate (\mbox{$T=80$ ns}), the experimental results (blue) follow the exact solutions up to $N\gtrsim100$, and the deviations are explained by the master equation solutions (black) using our noise model 
\cite{Note4}.
In comparison, using a sequence of standard gates, the experimental results (cyan) track the exact solutions only up to $N\sim10$, which fails to capture even a single period of the nonlinear oscillation.  
However, it is worth emphasizing that the hardware performance is comparable, and both quantum processors can perform $\sim 100$ gates with high fidelity. The difference in results mostly comes from how the unitary operator is realized: Using the standard compilation approach, each $U$ is realized by $\sim20$ native gates, whereas using the modular compilation approach, $U$ is realized as a single gate. By reducing the gate depth, the achievable simulation depth is thus significantly increased, which in this case qualitatively changes whether a useful Hamiltonian simulation is feasible or not.

\begin{figure}[b]
	\begin{center}
		\includegraphics[width=0.28\textwidth]{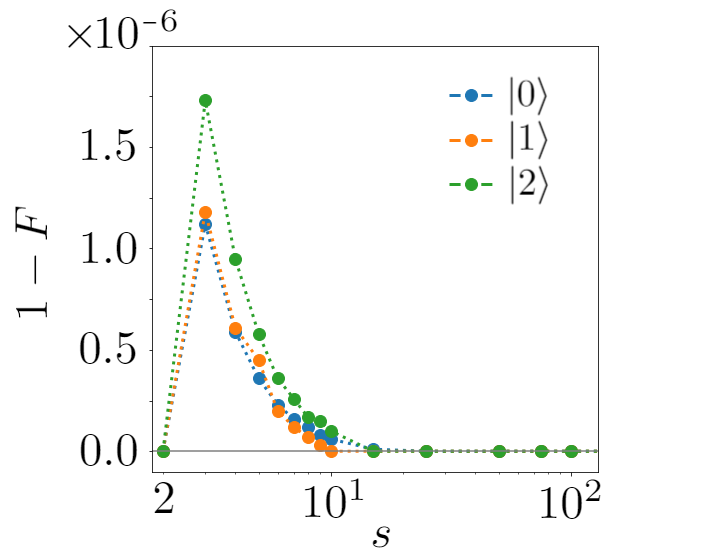} 
		\caption{Interpolated control pulses drive high-fidelity cubic gates for a wide range of $s$ parameters with the qudit initialized in $|0\rangle$, $|1\rangle$, and $|2\rangle$ states. The interpolation only uses optimized control pulses at $s=2$ and $s=\infty$ at fixed $\tau\sqrt{2s}=0.2$ and $\theta=\pi/2$.}
		\label{fig:Interpolation}
	\end{center}
\end{figure}

While offering better performance, modular gates are not more costly to compile on classical computers. 
The standard compilation approach first prepares a universal set of gates that are executable on quantum hardware, and then approximates a targeted $D\times D$ unitary operator $U$ using these native gates. Apart from the cost of preparing the gate set, compiling an arbitrary $U$ requires $O(D^2)=O(2^{2n})$ native gates \cite{Nielsen2002quantum}. Exponential speedup is possible only when $U=\exp(-iHt)$ and $H$ is native to the hardware.
When $H$ is not native, sophisticated quantum algorithms \cite{Childs2003exponential,Sgedy2004quantum,Childs2013universal,Berry2015hamiltonian,Low2017optimal} still have favorable complexity scaling, but are not readily implementable on hardware.
In comparison, directly compiling $U$ as a modular gate using optimal control requires $O(D^2)$ classical steps as well. However, once the control pulse is obtained, the unitary operator is enacted much more efficiently as a single pulse. 
Then, compared with classical computers, which always take $O(D^2)$ operations to compute $U\psi$, quantum processors may be used as special-purpose hardware accelerators for matrix multiplications. 
Moreover, the overhead of modular compilation may be amortized when the same $U$ is needed repeatedly or when a family of $U$ can be compiled without expensive numerical optimization.

In fact, high-fidelity parametric gates may be compiled cheaply using interpolation. 
For example, consider a one-parameter family of the cubic Hamiltonian 
$h(s)=\sqrt{2s}[(1-\xi)K(2) +(\xi-1/\sqrt{2})K(\infty)]/(1-1/\sqrt{2})$, where $\xi(s)=\sqrt{1-1/s}$, and the nonzero elements of the symmetric $3\times 3$ matrix $K$ are $K_{12}(s)=\xi$ and $K_{23}(s)=1$.
Motivated by this form, the interpolated control pulse for $s\ge 2$ at each time slice is taken to be $\epsilon_{\text{I}}(s)=[(1-\xi)\epsilon_{\text{O}}(2) +(\xi-1/\sqrt{2})\epsilon_{\text{O}}(\infty)]/(1-1/\sqrt{2})$, where $\epsilon_{\text{O}}(s)$ denotes the optimized pulse and $\tau\sqrt{2s}=0.2$ is held a constant. In other words, we interpolate the control pulse using the same formula for interpolating the Hamiltonian.   
To test this scheme, we numerically solve the master equation to obtain the gate fidelity, which is defined by \cite{Nielsen2002quantum} $F(\rho, \sigma)=\text{tr}\sqrt{\rho^{1/2}\sigma\rho^{1/2}}$, where $\rho$ and $\sigma$ are the density matrices attained using optimized and interpolated pulses after one gate application.
To optimize control pulses for $\rho(s)$, $\epsilon_{\text{O}}(2)$ is used as the initial guess. 
Although this compilation shortcut is not proven to work in general, the interpolated pulses are able to drive high-fidelity results for the three-level problem (Fig.~\ref{fig:Interpolation}).

In summary, we demonstrate that 
nonnative cubic interactions can be programmed entirely at the software level using an efficient action-space algorithm. 
Moreover, modular compilation reduces the requisite gate depth and qualitatively improves performance on noisy hardware. 
The resultant high-fidelity gates provide necessary building blocks for Hamiltonian simulations of a large class of nonlinear problems.

\begin{acknowledgements}
We would like to thank Rigetti Computing for providing access to their 16Q Aspen-4 processor, where we used lattice Aspen-4-2Q-A. 
We thank Max D. Porter for verifying numerical results in Figs.~\ref{fig:Control} and \ref{fig:Results}.
Y. S. would like to thank Eric T. Holland and Hong Qin for helpful discussions. 
This work was performed under the auspices of US Department of Energy (DOE) by LLNL under Contract DE-AC52-07NA27344. The experimental work was supported by the DOE Office of Fusion Energy Sciences ``Quantum Leap for Fusion Energy Sciences'' under project FWP-SCW1680, and the theory work was supported by LLNL-LDRD under Project 19-FS-072. 
The QuDIT hardware was funded under the National Nuclear Security Administration (NNSA) Advanced Simulation and Computing (ASC) ``Beyond Moore's Law'' quantum program under NA-ASC-127R-16 and US DOE, Office of Science, Office of Advanced Scientific Computing Research, Quantum Testbed Pathfinder Program under Award 2017-LLNL-SCW1631.
Y. S. was supported by the Lawrence Fellowship through LLNL-LDRD under Project 19-ERD-038.
\end{acknowledgements}


%

\onecolumngrid
\appendix

\renewcommand{\thefigure}{S\arabic{figure}}
\setcounter{figure}{0}  

\renewcommand{\theequation}{S.\arabic{equation}}
\setcounter{equation}{0}  

\clearpage
	\section{Unitary evolution of three levels}
For the three-level problem, the Hamiltonian matrix [Eq.~(11)] can be exponentiated analytically. The Schr\"{o}dinger time evolution of the three levels is determined by the unitary matrix $U=\exp(-ih\tau)$, which is given explicitly by
\begin{equation}
	U(\tau, \theta, s)=\left(\begin{array}{ccc} 
		\frac{(s-1)\cos\lambda\tau +s}{2s-1} & -ie^{i\theta}\sqrt{\frac{s-1}{2s-1}}\sin\lambda\tau & e^{2i\theta}\frac{\sqrt{s(s-1)}}{2s-1}(\cos\lambda\tau-1) \\
		-ie^{-i\theta}\sqrt{\frac{s-1}{2s-1}}\sin\lambda\tau  & \cos\lambda\tau & -ie^{i\theta}\sqrt{\frac{s}{2s-1}}\sin\lambda\tau \\
		e^{-2i\theta}\frac{\sqrt{s(s-1)}}{2s-1}(\cos\lambda\tau-1)  & -ie^{-i\theta}\sqrt{\frac{s}{2s-1}}\sin\lambda\tau & \frac{s\cos\lambda\tau +s-1}{2s-1}
	\end{array} \right),
\end{equation}
where $\lambda=\sqrt{2(2s-1)}$ is the positive eigenvalue of the Hamiltonian. The above $3 \times 3$ unitary matrix can be embedded into a two-qubits system by acting it on the $|00\rangle$, $|01\rangle$, and $|10\rangle$ states, while leaving the $|11\rangle$ state invariant.

The above unitary matrix is an input for both the standard and the modular compilation approaches. 
Although it may seem that once the unitary matrix is known, the problem would have already been solved, it is worth noting that the cubic gate may be a subproblem of some more complex problems. For example, in radiation hydrodynamics, lasers couple via three-wave interactions for given plasma conditions, while plasma conditions evolve due to laser energy deposition. The self consistent equations may be solved using a splitting algorithm, whose solutions may be written as $U_N V_N\dots U_1V_1U_0V_0$, where $U$'s are cubic gates and $V$'s are gates that advance the plasma conditions. In problems of this type, knowing $U$ is an intermediate step towards solving the entire problem, and quantum computing may be used to effectively carry out the matrix multiplications by applying precompiled gates.  

\section{Realizing cubic gates on Rigetti QCS}
\begin{figure}[t]
	\begin{center}
		\includegraphics[width=0.6\textwidth]{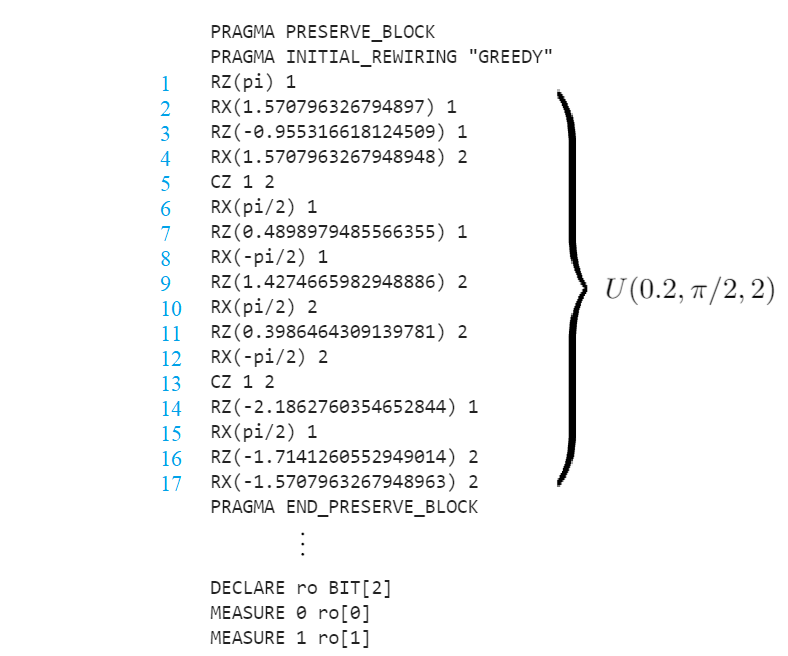} 
		\caption{An example Quil program that implements $U(0.2, \pi/2,2)$ on Rigetti's Aspen-4. In this example, the cubic gate is approximated by 17 native gates. This gate sequence can be repeated $N$ times to realize $U^N$ before the final readout.}
		\label{fig:Rigetti}
	\end{center}
\end{figure}

As an example of state-of-the-art quantum cloud services (QCS), Rigetti Computing \cite{Karalekas2020quantum,Smith2016practical} offers a native gate set consists of single-qubit rotations $\textsf{R}_x(\theta), \textsf{R}_z(\theta)$, and two-qubit \textsf{CZ} gates. The Quil compiler uses these gates to approximate other unitary operators using non-deterministic algorithms \cite{Smith2016practical}, which efficiently generate approximations for a given error tolerance. In particular, the unitary matrix $U(\Delta\tau, \theta, s)$ can be declared as a user-defined gate via the \texttt{DEFGATE} directive in the pyQuil library \cite{Smith2016practical}. 
After routing to two adjacent qubits on the quantum hardware, the probabilistic compiler typically converts this cubic gate to a sequence of $\sim 20$ native gates, including two \textsf{CZ} gates (Fig.~\ref{fig:Rigetti}). When directly repeating the gate in pyQuil, the compiler multiplies $[U(\Delta\tau)]^N=U(N\Delta\tau)$ and compiles for $U(N\Delta\tau)$ instead, so the hardware performance is independent of $N$.
This default simplification, which offloads the burden of computation to classical computers, is suppressed by placing the gate sequence for $U(\Delta\tau)$ within \texttt{PRAGMA PRESERVE\_BLOCK} and \texttt{PRAGMA END\_PRESERVE\_BLOCK}. In this way, $[U(\Delta\tau)]^N$ is realized on Rigetti's Aspen-4-2Q-A by applying a precompiled $U(\Delta\tau)$ for $N$ times, and the results 
\footnote{Rigetti QCS is continuously upgraded and routinely calibrated. For data reported in the main text, the QCS was accessed on January 7th, 2020 at 12:30 PM.}  
are shown in Fig.~2 of the main text.

\section{Noise modeling using Lindblad master equation}
Realistic quantum computers are open systems, and coupling to the environment is inevitable especially during control and readout operations.
The state of a quantum processor, which is mixed with the environment, may be characterized by its density matrix. Assuming processes in the environment are stationary and Markovian, then the time evolution of the density matrix may be described by the Lindblad master equation \cite{Louisell1973quantum,Lindblad1976generators,Gorini1976completely,Schlosshauer2019quantum}
\begin{equation}
	\partial_t\rho = i[\rho, H_0+H_c(t)] + \sum_j\big(L_j\rho L_j^\dagger -\frac{1}{2}\{L_j^\dagger L_j, \rho\}\big),
\end{equation}
where we have used the unit $\hbar=1$.
The first term is the unitary evolution due to the bare Hamiltonian $H_0$ of the quantum hardware, as well as the control Hamiltonian $H_c(t)$ due to the application of the control pulse. In the second term, Lindblad operators $L_j$ are used to model dissipative processes due to couplings with the environment. 
The above Lindblad master equation is solved numerically using the built-in function \textsf{mesolve} in QuTIP \cite{Johansson13, Johansson2012qutip}. 

To determine our hardware-specific Hamiltonians and Lindbladians, the transmon qudit is modeled using the Cooper pair box model whose parameters are measured experimentally \cite{Wu20PRL}. The measurements are made for the lowest three levels of the qudit and then extrapolated to higher levels.
First, the transition frequencies and the effective drive Hamiltonian are measured using Rabi spectroscopy. In the lab frame and in the transmon eigenbasis, keeping the lowest five levels, we approximate
\begin{equation}
	H_0\simeq\left(\begin{array}{ccccc} 
		0 & 0 & 0 & 0 & 0 \\
		0 & 25.758 & 0 & 0 & 0 \\
		0 & 0 & 50.099 & 0 & 0 \\
		0 & 0 & 0 & 72.848 & 0 \\
		0 & 0 & 0 & 0 & 93.828
	\end{array} \right), 
	\quad
	c\simeq\left(\begin{array}{ccccc} 
		0 & 1.000 & 0 & 0 & 0 \\
		0 & 0 & -1.372 & 0 & 0 \\
		0 & 0 & 0 & -1.618 & 0 \\
		0 & 0 & 0 & 0 & 1.781 \\
		0 & 0 & 0 & 0 & 0
	\end{array} \right), 
\end{equation}
where angular frequencies are in units of rad/ns. 
Second, two Lindbladians are included to model decay and dephasing. 
The Lindblad operator $L_1\sim a$ describes successive decays to lower levels, whose nonzero matrix elements are $L_1(j,j+1)=1/\sqrt{T_1(j+1, j)}$. 
The Lindblad operator $L_2\sim a^\dagger a$ describes dephasing with respect to lower levels, whose nonzero matrix elements are $L_2(j,j)=1/\sqrt{T_2^*(j, j-1)}$.
The $T_1$ decay time is measured by readout delays, and the $T_2^*$ dephasing time is measured using Ramsey spectroscopy. Keeping the lowest five levels, we approximate
\begin{equation}
	L_1\simeq\left(\begin{array}{ccccc} 
		0 & 0.004 & 0 & 0 & 0\\
		0 & 0 & 0.006 & 0 & 0\\
		0 & 0 & 0 & 0.007 & 0\\
		0 & 0 & 0 & 0 & 0.009\\
		0 & 0 & 0 & 0 & 0
	\end{array} \right),
	\quad
	L_2 \simeq \left(\begin{array}{ccccc} 
		0 & 0 & 0 & 0 & 0 \\
		0 & 0.005 & 0 & 0 & 0 \\
		0 & 0 & 0.014 & 0 & 0 \\
		0 & 0 & 0 & 0.045 & 0 \\
		0 & 0 & 0 & 0 & 0.000
	\end{array} \right). 
\end{equation}
Parameters in the Hamiltonians and the Lindbladians may drift over time and change after each cool down. The above parameters were obtained from the Quantum Design and Integration Testbed (QuDIT) calibration that was performed immediately before the experimental runs that produced the data reported in the main text.

\end{document}